\documentclass[aps,prl,reprint,twocolumn,superscriptaddress,longbibliography]{revtex4-2}
\usepackage{amsmath}
\usepackage{amsfonts}
\usepackage{hyperref}
\hypersetup{
	colorlinks=true,
	linkcolor=blue,
	filecolor=blue,
	citecolor = blue,      
	urlcolor=cyan,
}

\usepackage{braket}
\usepackage{graphicx}
\usepackage{siunitx}

\newcommand{\sect}[1]{{\bf #1.--} }

\newcommand{\figref}[1]{Fig.~\ref{#1}}

\begin{document}

\title{Entanglement Entropy in Quantum Networks with Tunable Geometry}

\author{Andrea Azzali}
\affiliation{Technical University of Munich, TUM School of Natural Sciences, Physics Department, James-Franck-Str.\ 1, 85748 Garching, Germany}
\affiliation{Munich Center for Quantum Science and Technology (MCQST), Schellingstr. 4, München 80799, Germany}
\affiliation{Scuola Normale Superiore, Piazza dei Cavalieri 7, 56126 Pisa, Italy}
\affiliation{Department of Physics, University of Pisa, Largo B. Pontecorvo 3, 56127 Pisa, Italy}
\author{Sridevi Kuriyattil}
\affiliation{Clarendon Laboratory, University of Oxford, Parks Road, Oxford OX1 3PU, United Kingdom}
\author{Andrew J. Daley}
\affiliation{Clarendon Laboratory, University of Oxford, Parks Road, Oxford OX1 3PU, United Kingdom}
\author{Marilù Chiofalo}
\affiliation{Department of Physics, University of Pisa, Largo B. Pontecorvo 3, 56127 Pisa, Italy}
\affiliation{INFN, Sezione di Pisa, Largo B. Pontecorvo 3, 56127 Pisa, Italy}

\begin{abstract}
Tunable-range random graphs provide a versatile setting for exploring how geometry shapes quantum dynamics, with potential applications to analog quantum simulation, amorphous materials, and quantum information processing. Here, we investigate transport and entanglement growth in quantum many-body systems whose two-body interactions are encoded by such graphs, with sites as nodes and two-site couplings as edges. 
In particular, we study dynamics for quantum hopping on a random graph model with an effective range parameter. By tuning it, we interpolate between nearest-neighbor and all-to-all graphs, identifying a new localization regime that is robust in both single-particle and finite-density cases.
Our results establish tunable-range random graphs as a platform for studying the interplay between connectivity, transport, and entanglement in quantum many-body dynamics.
\end{abstract}
\maketitle

\sect{Introduction}
In recent years, there has been growing interest in quantum many-body systems whose interaction terms extend beyond nearest-neighbors. A prominent example is provided by all-to-all interactions with power-law decay, which have been experimentally investigated with trapped ions and dipolar gases~\cite{joshi_quantum_2020,yan_observation_2013,baier_extended_2016}, where tunable-range interactions can be realized. These advances have triggered extensive theoretical and numerical works~\cite{buyskikh_entanglement_2016,defenu_long-range_2023,defenu_out--equilibrium_2024}, predicting violations of Lieb–Robinson bounds and rich phase diagrams with transitions between short- and long-range regimes.

More generally, the structure of pairwise interactions in a quantum many-body system can be conveniently encoded in a graph, where nodes represent sites and edges denote two-body couplings~\cite{bentsen_fast_2019,bentsen_treelike_2019,hashizume_tunable_2022,kuriyattil_onset_2023}. This approach enables a systematic study of a fundamental question: how quantum dynamics -- particularly, entanglement growth and the emergence of universal behavior -- is affected by the structure of pairwise couplings, expressed in terms of the graph geometry. We note that such exotic graphs are realizable through moving atoms in a Rydberg tweezer array~\cite{hashizume_deterministic_2021,bluvstein_logical_2024}.

A key advantage of this perspective is that it enables the description of complex interaction patterns using graph- and network-theoretic tools, which are being increasingly leveraged in quantum many-body problems~\cite{valdez_quantifying_2017,nokkala_reconfigurable_2018,nokkala_complex_2024,bhakuni_diagnosing_2024,ausilio_memory_2026}.
Rather than focusing on a specific graph configuration, it is convenient to define random graph models in terms of a few parameters, and characterize the resulting features in a probabilistic fashion. This generalization from deterministic to random graphs facilitates the identification of universal behaviors and encompasses a wide range of systems in which entanglement growth is of both theoretical and practical interest. Among these are disordered models, which may be used to investigate Anderson localization induced by structural disorder~\cite{de_luca_anderson_2014,sierant_universality_2023,cugliandolo_multifractal_2024,bhattacharjee_anderson_2025}, as well as many-body localization~\cite{tikhonov_anderson_2021} and glassy dynamics~\cite{lunkin_hilbert_2026}, and have also been studied experimentally in the context of amorphous materials~\cite{toh_synthesis_2020,tian_disorder-tuned_2023}. Furthermore, these models can be employed as resources for quantum information processing, such as substrates for Quantum Reservoir Computing~\cite{fujii_harnessing_2017,mujal_opportunities_2021}. Recent approaches have also linked the computational performance of quantum reservoirs to physical properties, including quasiparticle excitations~\cite{kobayashi_quantum_2025-1} and criticality~\cite{kobayashi_quantum_2025}. This perspective suggests that the graph structure itself may provide a means to tune the information-processing properties of the reservoir, by controlling the dynamical regime and the channels through which information propagates.

So far, the problem of quantum dynamics on graphs has been addressed in the case of the Bethe lattice~\cite{mirlin_localization_1991,baroni_corrections_2024} and subsequently in several random graph models~\cite{cugliandolo_multifractal_2024,bhattacharjee_anderson_2025},
with particular attention devoted to the ergodic or multifractal nature of delocalized states~\cite{mirlin_transition_1996,mirlin_exact_2006,de_luca_anderson_2014,cugliandolo_multifractal_2024}.
By contrast, tunable-range random graphs, in which the probability of forming links between nodes scales as a power-law with their distance, remain largely unexplored from the perspective of quantum dynamics, having been studied mainly from a graph-theoretic viewpoint~\cite{aizenman_discontinuity_1986,leuzzi_dilute_2008,brezin_crossover_2014,gori_one-dimensional_2017,millan_complex_2021}.

We aim to bridge this gap, giving insights into the role played by effective geometry in determining their dynamical properties.
To do so, we consider a random graph model~\cite{gori_one-dimensional_2017} that encompasses both the short- and long-range regime, and a Hamiltonian only consisting of hopping terms along the edges of the graphs, with the nodes corresponding to qubits. By choosing this paradigmatic Hamiltonian, we isolate the role of graph geometry in the dynamics, avoiding additional model-specific complications.
Remarkably, tuning from the short-range to the long-range regime, we not only recover the known behavior of the benchmarks -- namely, the nearest-neighbor and all-to-all models --, but also find an intermediate region with unexpected localization.
We then explain the new localized regime in terms of disorder in the random geometry of the graphs.

\begin{figure}
    \centering
    \includegraphics[width=0.99\linewidth]{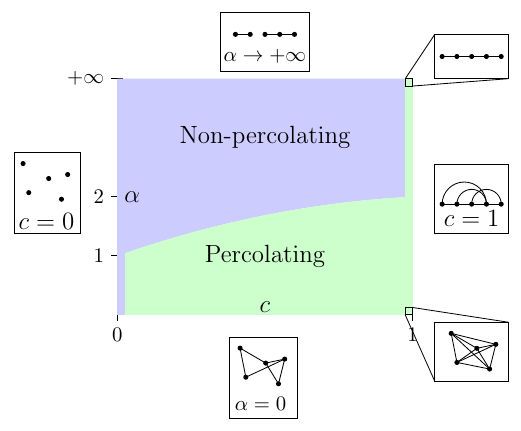}
    \caption{Schematic view of the classical phase diagram~\cite{gori_one-dimensional_2017} for the graph model in Eq. \eqref{eq:link-prob} in the space of the two parameters $(c,\alpha)$. The non-percolating region comprises the upper area of the plot and the $c=0$ line (here enlarged for clarity). The percolating region includes the lower area and the $c=1$ line (also enlarged). The insets show typical graphs for specific values of the parameters corresponding to the following benchmarks: totally disconnected ($c = 0$); nearest-neighbors connected ($c=1$), including the 1D chain ($\alpha \rightarrow +\infty$) and the fully-connected graph ($\alpha = 0$); the Erdős--Rényi model ($\alpha = 0$); the 1D chain with missing links ($\alpha \rightarrow +\infty$). \\
    It should be noticed that, tuning the parameters, a great variety of graphs can be explored. Decreasing $c$ amounts to randomly deleting some links, and changing $\alpha$ affects the probability of forming $r \geq 2$ links.}
    \label{fig:classical-phase-diagram}
\end{figure}

\sect{Model}
We consider a model consisting of a random graph and of a quantum Hamiltonian built on top of it.
The probability distribution on the graphs is taken from~\cite{gori_one-dimensional_2017}, and is defined in terms of two parameters, $c \in [0,1]$ and $\alpha \in [0,+\infty)$, as follows: In a graph of $N$ nodes, with integer labels $1, \ldots,N$, the probability that nodes $j$ and $\ell$ are linked is
\begin{equation}
    \frac{c}{|j - \ell|^\alpha} .
    \label{eq:link-prob}
\end{equation}
It is known~\cite{gori_one-dimensional_2017} that this model exhibits, for $\alpha \in (1,2)$, a percolation phase transition (see \figref{fig:classical-phase-diagram}). We now wish to explore the extent to which quantum effects can affect this phase diagram.

For every graph, we consider the dynamics of $N$ qubits, each corresponding to a node in the graph, with the Hamiltonian
\begin{equation}
    H = J \sum_{j,\ell} A_{j \ell} ~ \sigma^+_j \sigma^-_\ell ,
    \label{eq:ham-qubits}
\end{equation}
where $\sigma^{\pm} = (\sigma^x \pm i\sigma^y)/2$ are the spin-raising and -lowering operators, and $A$ is the adjacency matrix of the graph, whose entries $A_{j \ell}$ are $1$ if nodes $j$ and $\ell$ are connected, and $0$ otherwise. Since the adjacency matrix is symmetric and real, the Hamiltonian is Hermitian. \\
By introducing hard-core boson operators $b_j^\dagger = \sigma_j^-$ and $b_j = \sigma_j^+$, satisfying
\begin{align*}
    [b_j, b_\ell] = [b_j^\dagger, b_\ell^\dagger] = 0, \qquad [b_j, b_\ell^\dagger] = \delta_{j\ell} (\mathbb{I} - 2n_j),
\end{align*}
with $n_j = b_j^\dagger b_j$, Eq.~\eqref{eq:ham-qubits} can be rewritten as a random hopping Hamiltonian for hard-core bosons,
\begin{equation}
    H = J \sum_{j\ell} A_{j\ell} \, b_j^\dagger b_\ell .
    \label{eq:ham-hcbs}
\end{equation}

\sect{Methods}
We observe that the total number of hard-core bosons $N_\text{occ} = \sum_j n_j$ is conserved, and begin by studying dynamics in the $N_\text{occ} = 1$ sector, whose size grows linearly with $N$. We use Monte Carlo sampling to sample over the graphs distribution; within each graph, the edges are independently sampled.

\begin{figure}
    \centering
    \includegraphics[width=0.99\linewidth]{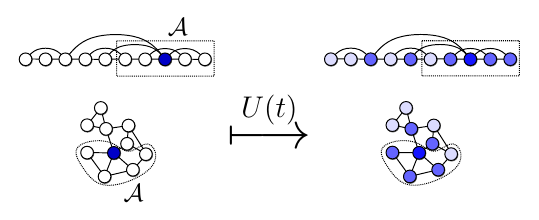}
    \caption{Schematic depiction of the simulation performed on each graph sample. The figure shows the system before (left) and after (right) unitary time evolution $U(t)$; the two snapshots are represented according to the Euclidean geometry (top) and highlighting the graph structure (bottom). For each graph sample, the simulation is performed in the following way: The system is divided into two halves, subsystem $\mathcal A$ and its complementary $\overline {\mathcal A}$, and is initialized with a single particle in $\mathcal A$. Then, unitary time evolution $U(t)$ induced by the Hamiltonian of \eqref{eq:ham-hcbs} acts on the system.
    This schematic figure highlights that the spreading is more easily understood in terms of the graph geometry, rather than the Euclidean one, as argued in the following sections.}
    \label{fig:setup}
\end{figure}

\textit{Single-particle case.} For each graph, we use Exact Diagonalization (ED) to determine the dynamics, starting from a state of the form $b^\dagger_j \ket{\Omega}$, where $\ket \Omega$ is the vacuum.
In practice, we compute the entanglement entropy $S = - \text{tr}[\rho_\mathcal A \log \rho_\mathcal A]$ for the subsystem $\mathcal A = [1,N/2] \cap \mathbb Z$, where $\rho_\mathcal A$ is the reduced density-matrix for $\mathcal A$, and the whole system is always in a pure state. Entanglement entropy is computed as a function of time, initially positioning the particle at $N/4$ (see \figref{fig:setup}). To check if the results are affected by closeness to the boundary, simulations have also been performed with $\mathcal A = [N/4,3N/4] \cap \mathbb Z$ and initially locating the particle at $N/2$, without any difference observed. For the single-particle sector, it should be noted that the entanglement entropy is the Shannon entropy of the probability vector $(p,1-p)$, with $p$ being the probability that the particle lies in $\mathcal A$.

\textit{Half-filling case.} The dimension of sectors with finite density grows exponentially with $N$, making ED unfeasible for large system sizes. For this reason, we replace the hard-core boson operators $b,b^\dagger$ in the Hamiltonian \eqref{eq:ham-hcbs} with operators $c,c^\dagger$, such that
\begin{equation*}
    \{ c_j,c_\ell \} = \{ c^\dagger_j,c^\dagger_\ell \} = 0 \qquad \{ c_j,c^\dagger_\ell \} = \delta_{j \ell} ,
\end{equation*}
corresponding to spinless fermions.
This replacement is equivalent to applying a Jordan--Wigner transformation and deleting the Jordan--Wigner strings. In any case, the new Hamiltonian is not equivalent to the previous one, except for the single-particle sector.
This new Hamiltonian, however, is quadratic in the fermionic operators and is therefore non-interacting. Exploiting fermionic Gaussian states, it is now possible to determine correlation functions and entanglement entropy at the cost of diagonalizing an $N \times N$ matrix~\cite{peschel_calculation_2003,surace_fermionic_2022}.
We numerically simulate the dynamics in the half-filling sector, i.e. $N_\text{occ} = N/2$, with the sites in $\mathcal A = [1,N/2] \cap \mathbb Z$ initially occupied. We compute the entanglement entropy for subsystem $\mathcal A$ and the expected number of particles outside $\mathcal A$, as functions of time.
Simulations have also been performed at other values of filling, yielding analogous results.

\sect{Results}
In all the numerical simulations we observe that, after averaging, the entropy displays an increase from the initial value of $0$, eventually saturating to a value which we denote as $S_\infty$ (see \figref{fig:example-S-vs-t}). The same applies to the number of particles outside the subsystem in the half-filling case, converging to $n_\infty$. Therefore we will mainly be interested in these asymptotic values, as a function of $c$, $\alpha$ and $N$, relating them to the localizing or delocalizing behavior of the system.

We begin by studying the case $c = 1$, in which the graphs are connected, as the probability of nearest-neighbor links is $1$. This is the simplest setup for studying quantum localization effects. Conversely, for $c < 1$ one might observe localization merely because the particle initially lies in a small cluster, in the non-percolating phase. We will revert to this less straightforward case at the end.

\begin{figure}
    \centering
    \includegraphics[width=0.9\linewidth]{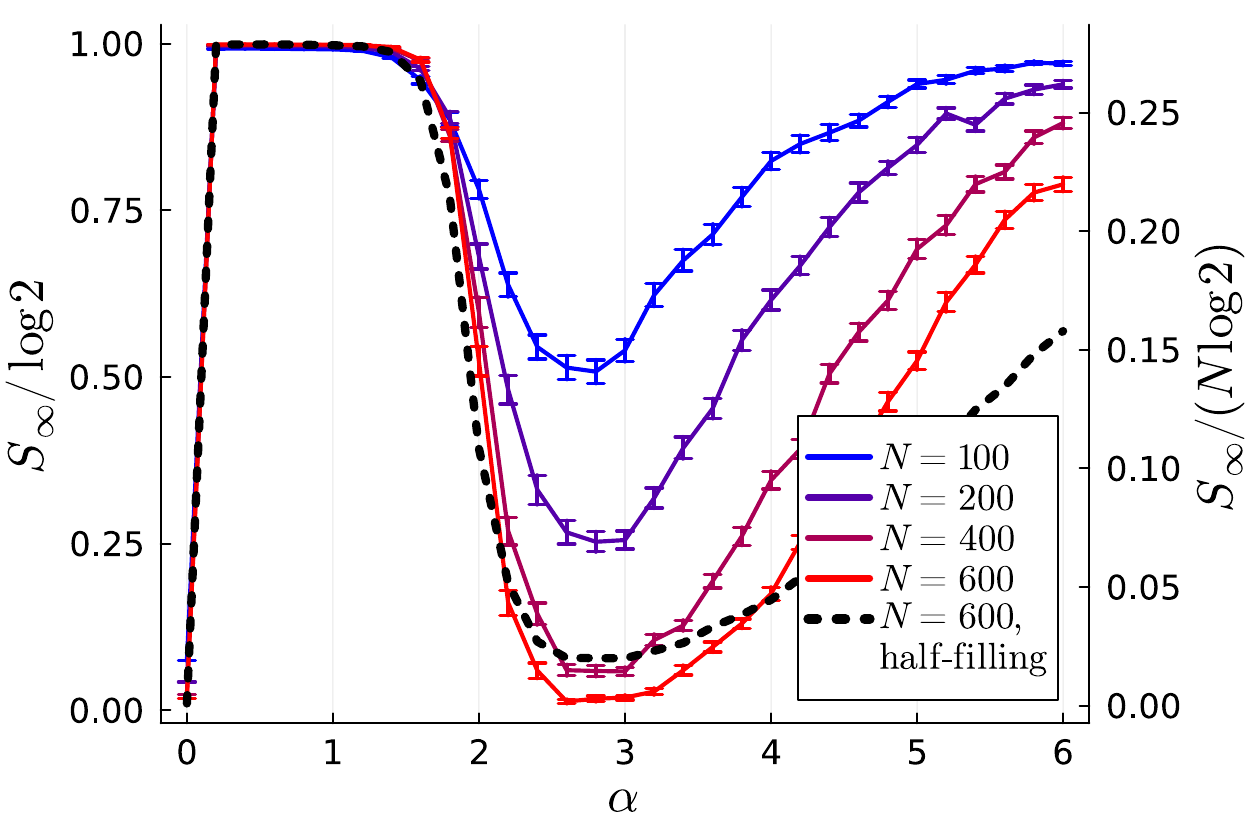}
    \caption{Asymptotic entanglement entropy $S_\infty$ as a function of $\alpha$ at $c=1$.
    This figure represents one main result: The asymptotic entanglement entropy exhibits a new localized regime at intermediate values of $\alpha$, while recovering the limiting behaviors at nearest-neighbors and all-to-all. This behavior is present both in the single-particle case (solid lines, left $y$ axis) and in the half-filling case (dashed line, right $y$ axis).}
    \label{fig:S-vs-alpha-single-particle}
\end{figure}

\textit{Case $c = 1$.} At long times, if the system is localized, we should observe $S_\infty \sim 0$, whereas with full delocalization $S_\infty$ should approach its maximal value $\log 2$.
The single-particle numerical simulations yield an unexpected result, shown in \figref{fig:S-vs-alpha-single-particle}. We recover the analytically known behavior of the benchmarks at $\alpha = 0$ (all-to-all model, fully localized as $N \rightarrow \infty$) and $\alpha \rightarrow +\infty$ (XX model, delocalized). As $\alpha$ decreases from $+\infty$ all the way down to 0, the probability of forming links is increased, which one would naively expect to enhance delocalization (i.e., increase $S_\infty$). However, this is contradicted by the non-monotonous behavior of the curve in \figref{fig:S-vs-alpha-single-particle}, which reaches maximally entangled values for $\alpha \lesssim 2$ and in the limit of $\alpha \rightarrow +\infty$, but shows an intermediate localized regime. This regime is also observed in the Inverse Participation Ratio (see Supplemental Material).

\begin{figure}
    \centering
    \begin{minipage}{0.9\linewidth}
        \includegraphics[width=0.9\linewidth]{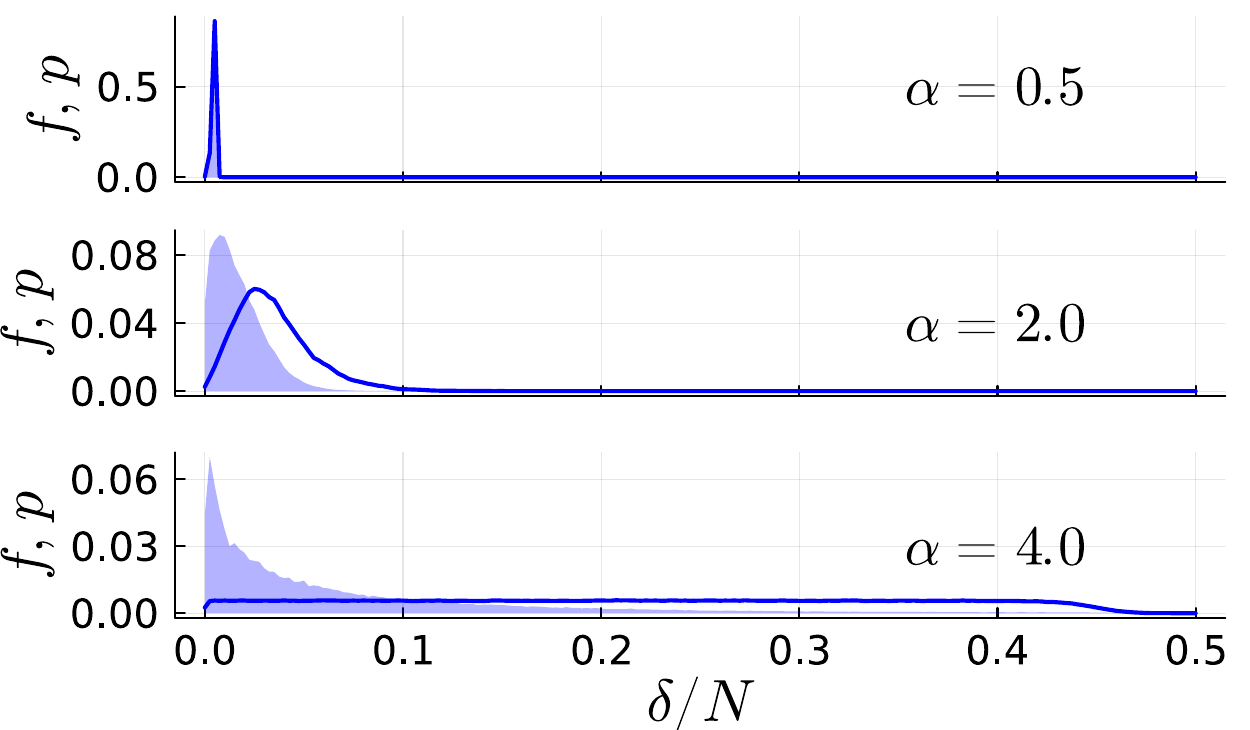}
    \end{minipage}
    \begin{minipage}{0.07\linewidth}
        (a) \vspace{0.9cm} \\ (b) \vspace{0.9cm} \\ (c) \vspace{0.5cm}
    \end{minipage} \\
    \vspace{0.2cm}
    \begin{minipage}{0.9\linewidth}
        \includegraphics[width=0.9\linewidth]{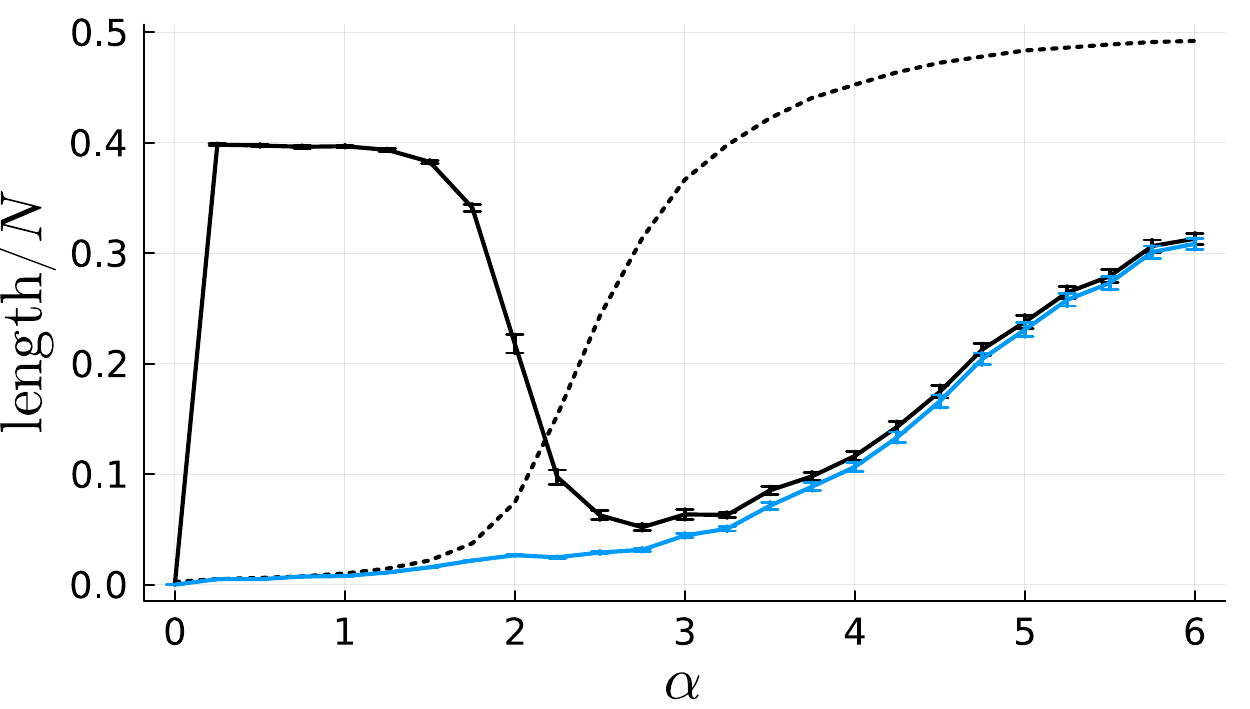}
    \end{minipage}
    (d)
    \caption{(a,b,c) Classical vs.\ quantum distributions for the graphs, with respect to $\alpha$. For different values of $\alpha$ in each subplot, we show the fraction $f(\delta)$ of nodes (solid lines) and the probability $p(\delta)$ at a graph distance $\delta$ (shaded regions).
    (d) Asymptotic wavefunction spreading as a function of $\alpha$. The curves represent the spreading length-scale in the Euclidean (black, solid line) and graph (light blue line) geometry. The dotted line is the maximum graph distance from the site located at $N/2$.
    All data in this figure refer to $N = 400$.}
    \label{fig:fraction-vs-graph-distance}
\end{figure}

To understand this minimum of $S_\infty$, we connect it to the graph geometry, defining the graph distance $d_G$ as the length of the shortest path between two nodes:
\begin{equation}
    d_G(j,\ell) = \min \{m \in \mathbb N ~|~ (A^m)_{j \ell} \neq 0 \},
    \label{eq:graph-distance}
\end{equation}
where $A$ is the adjacency matrix of the considered graph $G$.
\figref{fig:fraction-vs-graph-distance} shows two relevant quantities as a function of $\alpha$, averaged over graph samples: 
The first one is the distribution of nodes as a function of the graph distance $\delta$ from the node $N/2$
\begin{equation*}
    f(\delta) = \left \langle \frac 1 N ~ \#\{j ~ | ~ d_G(N/2,j) = \delta \} \right \rangle,
\end{equation*}
and the second is the probability for the particle to lie at a graph distance $\delta$
\begin{equation*}
    p(\delta) = \left \langle \sum_{d_G(N/2,j) = \delta} ~ |\psi(j)|^2 \right \rangle .
\end{equation*}
The distribution $f(\delta)$ is a statistical property of the graphs, whereas $p(\delta)$ is obtained by simulating quantum dynamics.

We now connect the behavior of $f(\delta)$ and $p(\delta)$ to the entanglement entropy as a function of $\alpha$ (\figref{fig:S-vs-alpha-single-particle}). For the smallest values of $\alpha$ (\figref{fig:fraction-vs-graph-distance}a), most nodes can be reached within few steps and the probability spreading is almost uniform on all sites, hence $S_\infty$ reaches the maximum. As $\alpha$ is increased (\figref{fig:fraction-vs-graph-distance}b,c), $f(\delta)$ becomes flatter, but the probability is still concentrated on the closest neighbors, displaying a localized behavior and the minimum observed in $S_\infty$.

This behavior is also observed in the asymptotic spreading of the wavefunction (\figref{fig:fraction-vs-graph-distance}d), computed at the large times also used to extract $S_\infty$. As a length-scale for this spreading, we consider the minimum distance from the initial particle position enclosing a fixed probability. The figure refers to a probability of $0.8$, but changing this value yields similar results. Two different curves are obtained considering the Euclidean and graph-induced distances; in the limit $\alpha \rightarrow +\infty$, the two distances become equivalent and delocalization is present. We note that the spreading length in the graph distance interpolates between this regime, for large $\alpha$, and the maximum graph distance, for the smallest values of $\alpha$. Instead, the spreading length in the Euclidean geometry has the same behavior of $S_\infty$ (see \figref{fig:S-vs-alpha-single-particle}) as a consequence of our choice of bipartition for the system.

In addition to this, we also gain insight on the behavior of \figref{fig:S-vs-alpha-single-particle} by looking at the large-$\alpha$ limit, in which the probability of forming a link of length $r > 1$ is suppressed as $r^{-\alpha}$. In this limit and at finite $N$, we approximate the system to a chain (the clean XX model) with some additional links of length 2. Treating these as impurities, and the whole system as disordered and one-dimensional, allows us to exploit the theory of Anderson localization~\cite{anderson_absence_1958,landauer_electrical_1970,anderson_new_1980}. From it, we obtain a localized wavefunction (see Supplemental Material for the details) and derive a prediction for $S_\infty(\alpha)$ that has been fit to the actual numerical data, with good agreement for large values of $\alpha$ (see \figref{fig:fit-scatterers-model}).
This picture also addresses the $N$ dependence visible in \figref{fig:S-vs-alpha-single-particle}: The typical length-scale of the localized wavefunction only depends on the density of impurities, hence on $\alpha$, and is independent of $N$. As a consequence, increasing the system size $N$ ensures that most of the probability density lies within subsystem $\mathcal A$, hence decreasing $S_\infty$.
The intuition we draw from this approximate model is that the unexpected localization is due to \textit{structural disorder}, i.e.\ to the non-homogeneity of the graph structure.

We now move on to the numerical simulations at half-filling. The behavior of entanglement entropy (see \figref{fig:S-vs-alpha-single-particle} and the Supplemental Material for details) is analogous to the single-particle case, but now the maximum value reached is $S_\infty / (N \log 2) \approx 0.28$. We note that this is compatible with the expected value in the thermodynamic limit if the system is in a random Gaussian state, $S/N \longrightarrow \log 2 - \frac 1 2$~\cite{bianchi_page_2021}. Moreover, the expected number of particles outside subsystem $\mathcal A$ converges to $N_\text{occ}/2$. Therefore, no memory of the initial state, apart from Gaussianity, is retained by the quantities $S_\infty$ and $n_\infty$.

\textit{Case $c < 1$.} If $c$ is decreased below $1$, there is a certain value of $\alpha = \alpha^*$ separating the non-percolating ($\alpha > \alpha^*$) and the percolating ($\alpha < \alpha^*$) regions of the phase diagram. This is a classical property, but it clearly affects quantum dynamics for our choice of initial states, as a single particle in a small cluster not coupled to the rest of the system cannot propagate. Hence, for $\alpha > \alpha^*$, we expect localization, whereas for $\alpha < \alpha^*$ one might or might not observe delocalization. Anyway, it should be noted that, for this model, $\alpha^* \in (1,2)$, while the unexpected localization in the $c=1$ case occurred for larger values of $\alpha$. In fact, the numerical results (see \figref{fig:S-vs-alpha-c-smaller-1}) seem to only exhibit a delocalization behavior for $\alpha < \alpha^*$, together with the expected localization for $\alpha > \alpha^*$, but larger system sizes would be needed to answer this question more accurately. At half-filling, the results at $c < 1$ are also analogous to the single-particle ones.

\sect{Conclusions and Outlook}
In this work, we have studied the role played by graph geometry in quantum hopping on a random graph model.
When the probability of nearest-neighbor links is $1$, we observe crossovers between a variety of regimes, determined by the interplay of disorder and interaction range. In particular, by tuning $\alpha$, we access a regime between Euclidean and all-to-all geometry that exhibits localization induced by the random structure of the graphs.
A hallmark of this localization is found in the entanglement entropy, both in the single-particle case and at finite density for free fermions.
Further signatures of this effect have been observed in the IPR and in the spectral properties of the adjacency matrix (see Supplemental Material), suggesting the presence of multifractality%
~\cite{mirlin_exact_2006,de_luca_anderson_2014,cugliandolo_multifractal_2024}.

Another interesting direction would be to replace our minimal model with random graphs tailored to describe specific amorphous materials~\cite{toh_synthesis_2020,tian_disorder-tuned_2023}: In this way, the robustness of this unexpected localization regime could be experimentally probed.

Furthermore, the different regimes of information spreading provide a route to engineering quantum reservoirs with controllable dynamical properties~\cite{fujii_harnessing_2017,mujal_opportunities_2021,de_lorenzis_entanglement_2026}. Conversely, systematically assessing the computational performance of tunable-range random graphs would give insight into the mechanisms that enable information propagation in these reservoirs~\cite{kobayashi_quantum_2025-1}.
This approach could be applied to the crossover at $\alpha \sim 2$, looking for an enhancement in performance similar to that usually observed at criticality~\cite{martinez-pena_dynamical_2021,kobayashi_quantum_2025}.

Future work could also explore non-Hermitian extensions of the setup, for instance with noise-engineering in the Markovian and non-Markovian regimes~\cite{luschen_signatures_2017,stanzione_tailoring_2025}.
Alternatively, one could consider non-reciprocal hopping, which on regular one-dimensional lattices is known to exhibit a localization transition~\cite{hatano_localization_1996,hatano_vortex_1997}, and may be extended to more complicated setups, also supporting the presence of topological invariants~\cite{rudner_topological_2009,yao_edge_2018,song_non-hermitian_2019}.

\sect{Acknowledgments}
A.A.\ and M.L.C.\ acknowledge useful discussion with Giuseppe La Rocca. A.A.\ acknowledges support by the Deutsche Forschungsgemeinschaft (DFG, German Research Foundation) under Germany’s Excellence Strategy – EXC-2111 – 390814868.
M.L.C.\ acknowledges support from the National Centre on HPC, Big Data and Quantum Computing—SPOKE 10 (Quantum Computing) and received funding from the European Union Next-GenerationEU—National Recovery and Resilience Plan (NRRP)—MISSION 4 COMPONENT 2, INVESTMENT N.\ 1.4—CUP N.\ I53C22000690001. This research has received funding from the European Union's Digital Europe Programme DIGIQ under Grant Agreement No.\ 101084035.
This work was supported by the EPSRC through the QQQS programme grant (EP/Y01510X/1) and by the QCi3 hub (EP/Z53318X/1).

\sect{Data availability}
Scripts and outputs of the numerical simulations are available upon reasonable request on Zenodo~\cite{azzali_entanglement_2026}.

\bibliography{TunableRangeQuantumNetworks,TunableRangeQuantumNetworks2,TunableRangeQuantumNetworks3}

\clearpage

\onecolumngrid
\begin{center}
\vskip0.5cm
{\Large Supplemental Material}
\end{center}
\vskip0.4cm

%%%%%%%%%% Prefix a "S" to all equations, figures, tables and reset the counter %%%%%%%%%%
\setcounter{section}{0}
\setcounter{equation}{0}
\setcounter{figure}{0}
\setcounter{table}{0}
\setcounter{page}{1}
\renewcommand{\theequation}{S\arabic{equation}}
\renewcommand{\thefigure}{S\arabic{figure}}
\renewcommand{\thesection}{S\arabic{section}}
\renewcommand{\thesection}{S\arabic{section}}
%%%%%%%%%%%%%%%%%%%%%%%%%%%%%%%%%%%%%%%%
%
\subsection{Convergence of entanglement entropy to its asymptotic value}
\begin{figure}[h]
    \centering
    \includegraphics[width=0.4\linewidth]{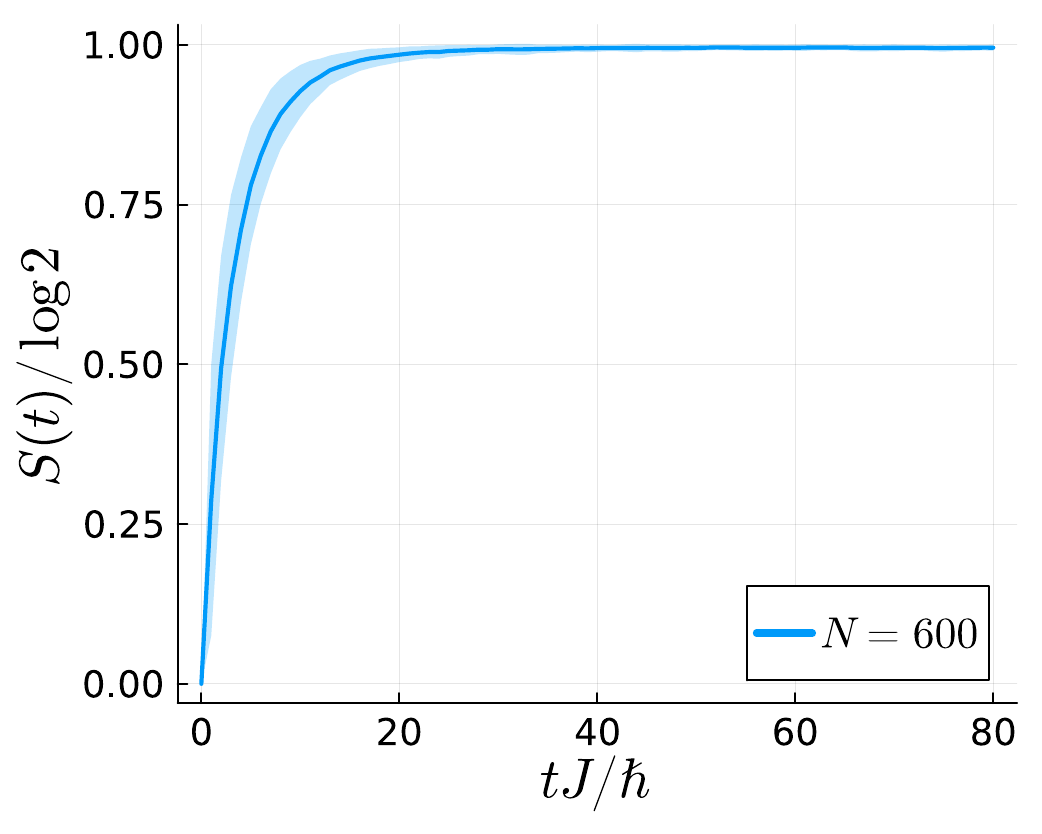}
    \caption{Example of average entanglement entropy $S$ as a function of time. The parameters are set to $N = 600$ sites, $c = 1$, $\alpha = 1.4$, $N_\text{occ} = 1$; $100$ graph samples are taken. The solid line is the average and the thickness of the colored shaded area is the standard deviation over the samples distribution. The identity $S(t = 0) = 0$ descends from the choice of the initial state.}
    \label{fig:example-S-vs-t}
\end{figure}

The asymptotic values are computed at time $t = 10^4 ~ \hbar/J$, which was checked to be a long enough time to reach saturation for all choices of the parameters used in the numerical simulations. \figref{fig:example-S-vs-t} shows the convergence of entanglement entropy to its asymptotic value $S_\infty$.

\subsection{Asymptotic value of the Inverse Participation Ratio (IPR)}
\newcommand{\ipr}{\text{IPR}}
A vastly used measure of delocalization is given by the Inverse Participation Ratio (IPR)~\cite{wegner_inverse_1980,mirlin_exact_2006,evers_anderson_2008}, corresponding to the second moment of the wavefunction's squared amplitude in the position basis:
\[ \ipr({\psi}) = \sum_j |\psi(j)|^{4} . \]
In the single-particle case, if the wavefunction is localized on a single site, then $\ipr \sim 1$, while if the amplitude is evenly spread across the sites, $\ipr \sim N^{-1}$, where $N$ is the number of sites.

\begin{figure}
    \centering
    \includegraphics[width=0.5\linewidth]{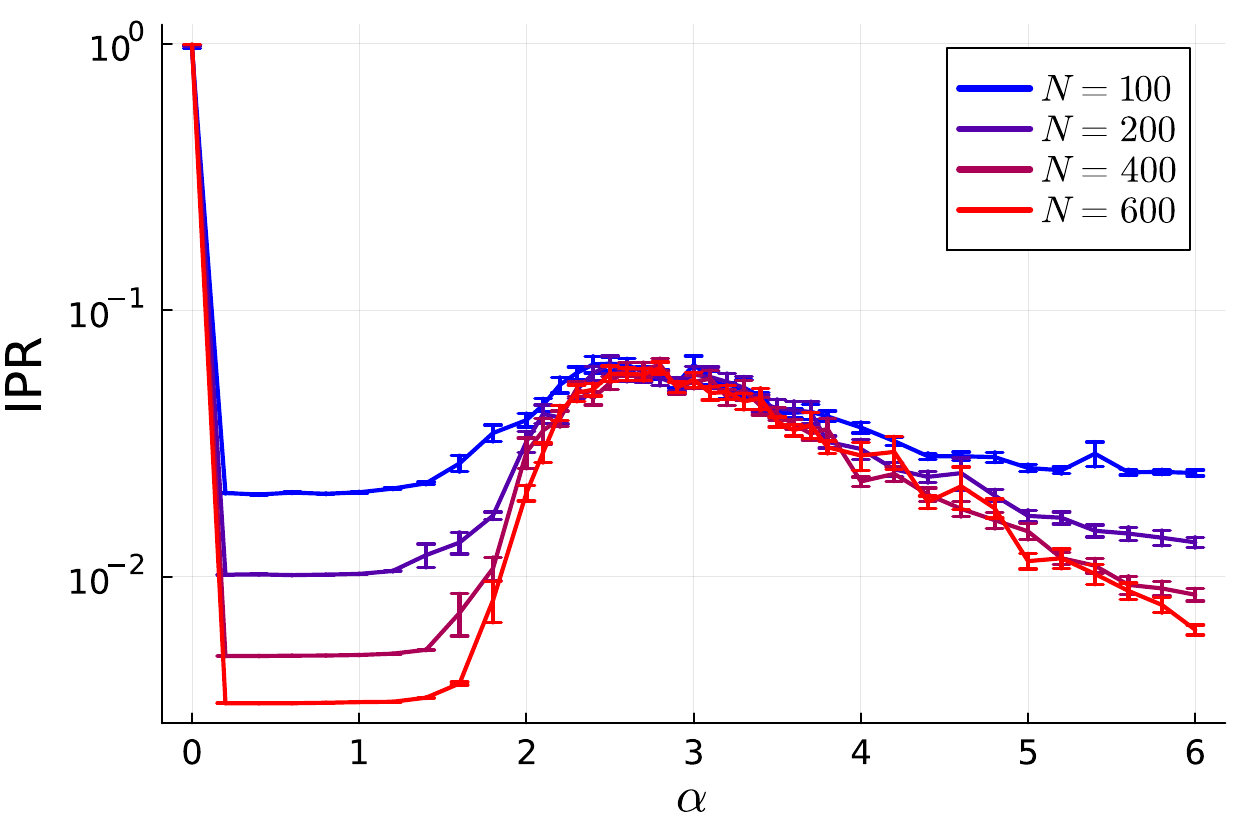}
    \caption{Asymptotic values of the IPR as a function of $\alpha$. The plot refers to the single-particle case with $c=1$, and values are averaged over $100$ graph samples. Again, we observe full localization at $\alpha = 0$, the localized regime around $\alpha \sim 3$ and the delocalized regime elsewhere.}
    \label{fig:ipr-from-dynamics}
\end{figure}

Here, we repeat the numerical simulations used above for the entropy, this time extracting the asymptotic value of the IPR from the dynamics. The outputs are shown in \figref{fig:ipr-from-dynamics}. The corresponding localization behavior is consistent with the one observed with the asymptotic entanglement entropy. This shows that our results for the entropy are not an artifact of the choice of bipartition, since the IPR does not involve such a bipartition. The value of the IPR is instead specific to the real-space basis, which however is a natural choice.

\subsection{Timescale for the saturation to asymptotic values}
\begin{figure}[h]
    \centering
    \includegraphics[width=0.5\linewidth]{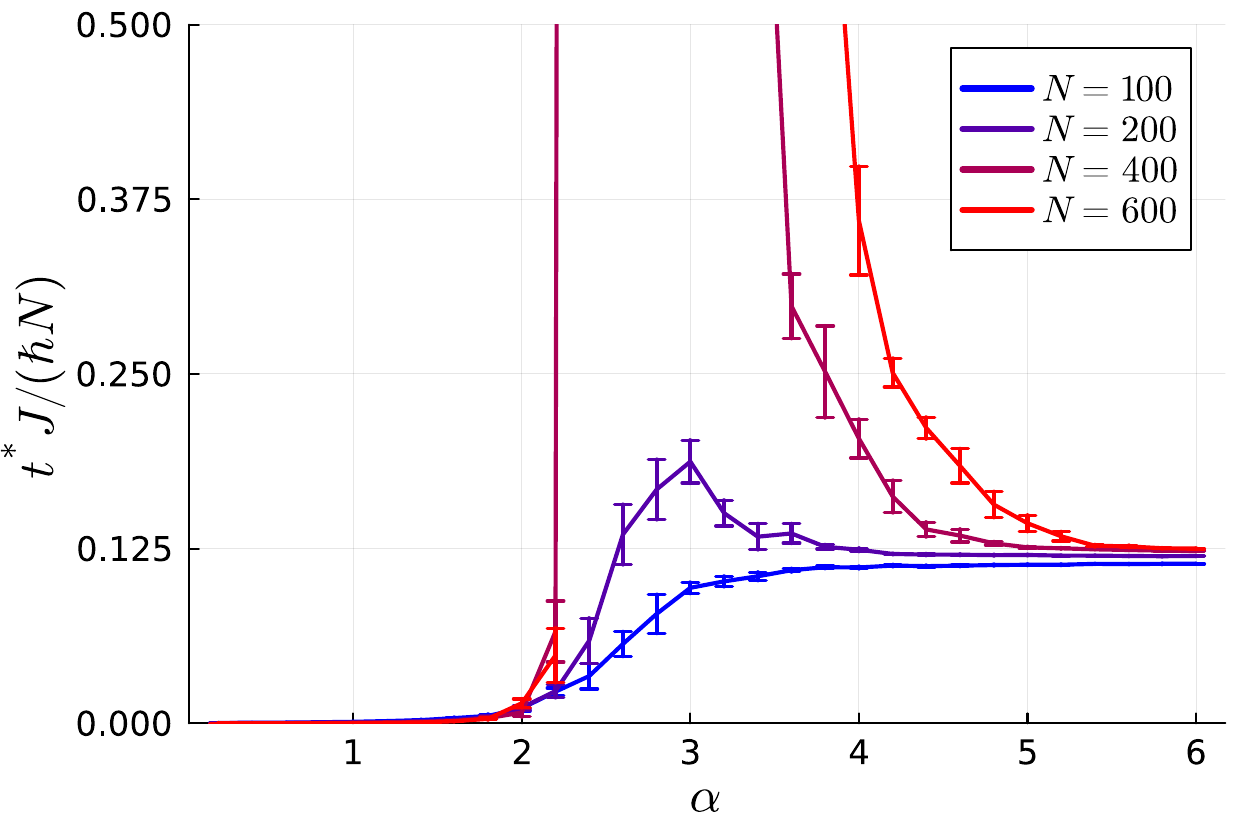}
    \caption{Timescale for the saturation of entanglement entropy to $S_\infty$ as a function of $\alpha$. The plot refers to the single-particle case with $c=1$ and a threshold entropy of $S^*/\log 2=0.1$. The divergence in $t^*$ is present when the asymptotic value is not larger than the threshold.}
    \label{fig:timescale}
\end{figure}

We numerically determine the timescale needed for entanglement entropy to saturate to its asymptotic value $S_\infty$, as a function of the system size $N$ and of the parameter $\alpha$.
Since the entropy, after averaging, is non-decreasing with time, we compute the time $t^*$ at which a threshold value of entropy $S^*$ is reached and use $t^*$ as an estimate of the saturation timescale (see \figref{fig:timescale}). It has been checked that changing the threshold does not affect the results significantly.

To understand the behavior of $t^*$ as a function of $\alpha$, we consider the spreading of entanglement to be described by entangled quasiparticle pairs~\cite{calabrese_evolution_2005,alba_entanglement_2018}.
We use as a benchmark the case of the XX model, with the single excitation initially located at $N/4$ and the boundary at $N/2$. Given that the system is one-dimensional, $t^*$ is linear in its size and, taking into account the speed of the fastest quasiparticle, $t^* J / \hbar \sim N/8$. For $\alpha \to + \infty$, we actually recover this behavior, as can be seen in \figref{fig:timescale}.
Conversely, for $\alpha \lesssim 2$, the timescale is extremely shorter because of the presence of long-range links that allow the quasiparticles to cross the bipartition.

\subsection{Results of the numerical simulations with $c < 1$}
\begin{figure}[h]
    \centering
    \includegraphics[width=0.4\linewidth]{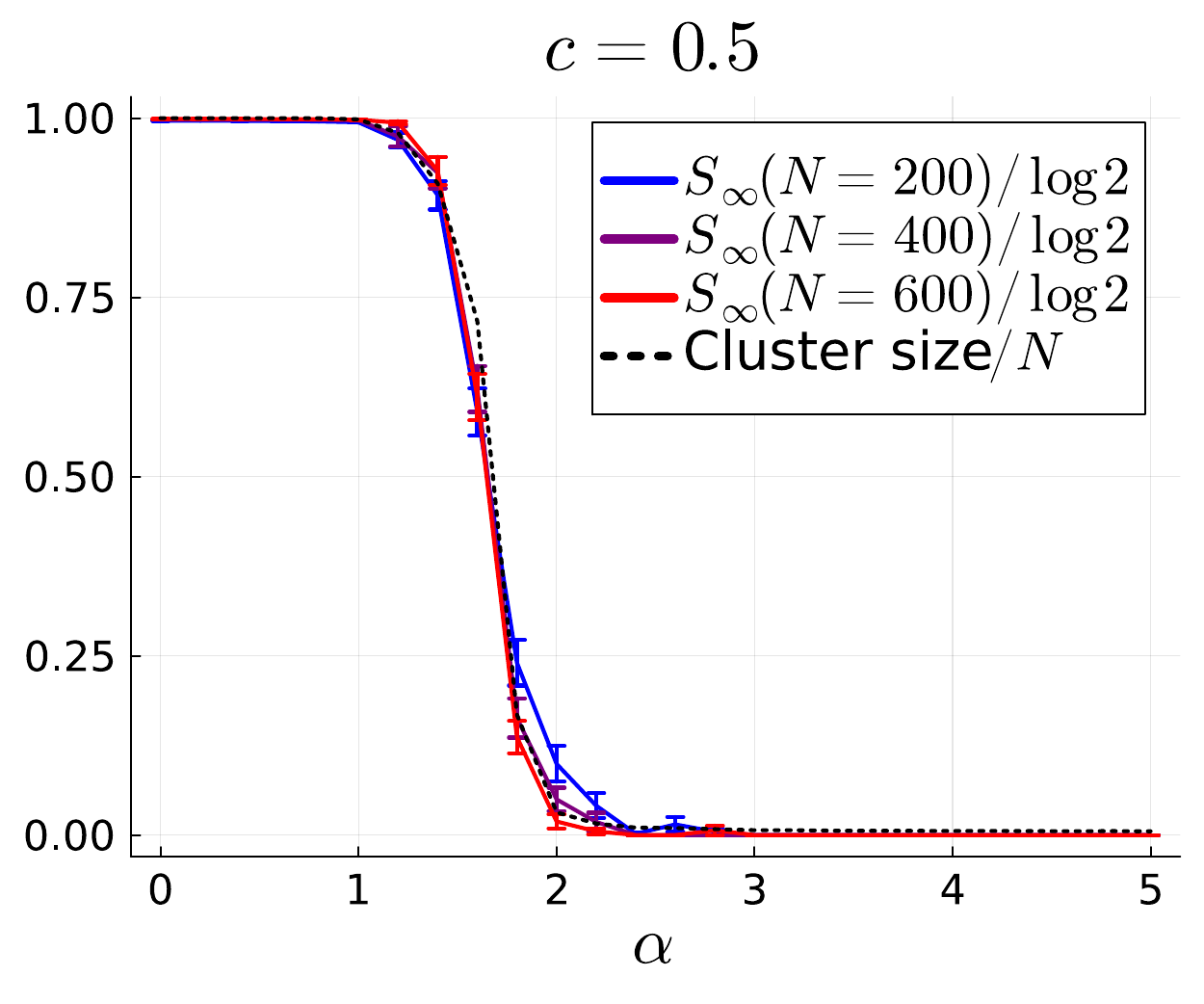}
    \includegraphics[width=0.4\linewidth]{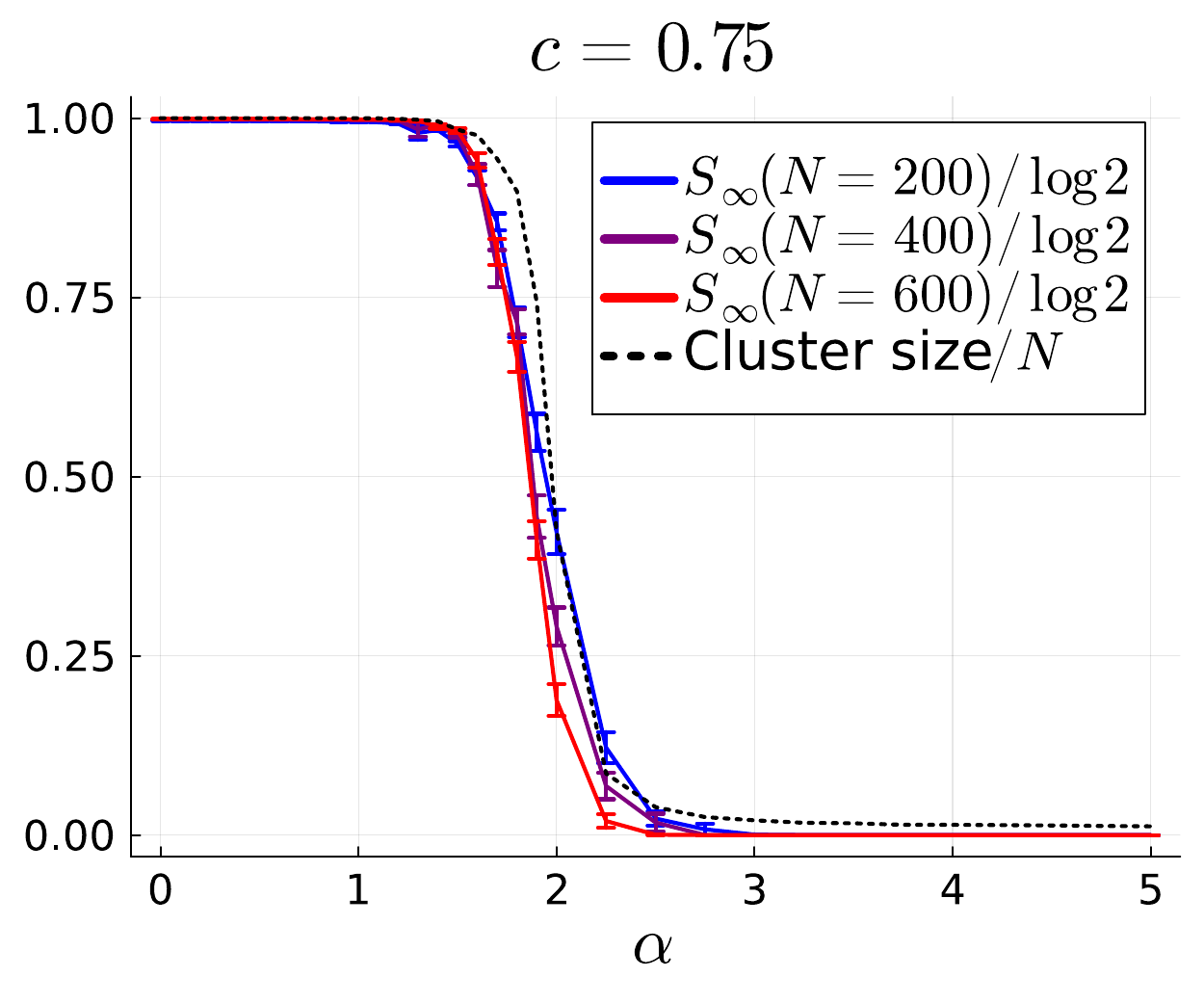}
    \caption{Classical vs. quantum behavior as a function of $\alpha$ for two exemplifying values of $c < 1$,  namely $c = 0.5$ (left panel) and $c = 0.75$ (right), in the single-particle case. Every value is averaged over $100$ graph samples. The colored solid lines represent the asymptotic entanglement entropy for several values of $N$ as in the legend. For comparison, the dotted black line shows the fraction of the sites within the connected component of the initial site: This measures the number of sites that can be reached from there, and is a property of the graphs. It is computed at the largest value of $N$ in the legend. Both the fraction of sites in the cluster and the entropy (rescaled by $\log 2$) range from $0$ to $1$, therefore sharing the vertical axis in these plots.}
    \label{fig:S-vs-alpha-c-smaller-1}
\end{figure}

\figref{fig:S-vs-alpha-c-smaller-1} shows the behavior of the asymptotic entanglement entropy in the case of $c < 1$ in the single-particle case. Simulations have also been performed at half-filling, yielding analogous results.

\subsection{Results of the numerical simulations in the half-filling case}
\begin{figure}[h]
    \centering
    \includegraphics[width=0.6\linewidth]{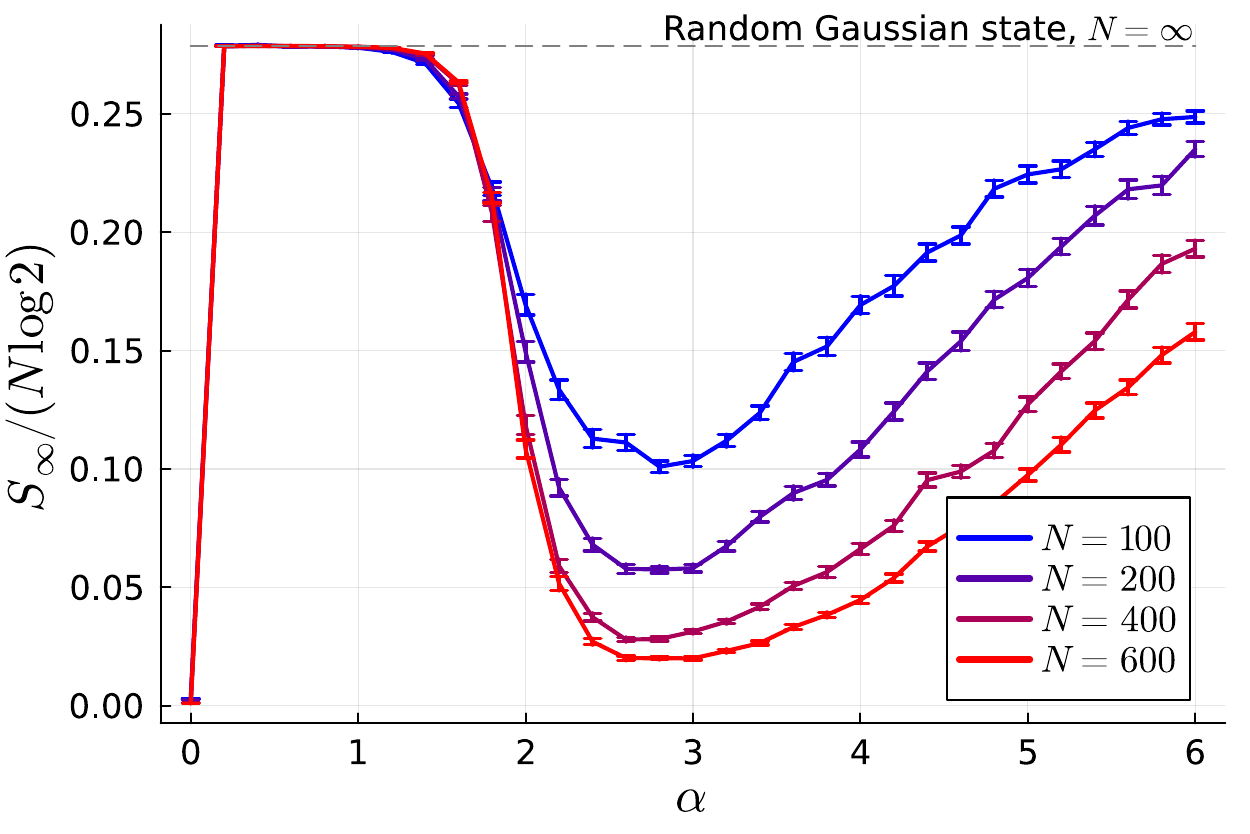}
    \caption{Asymptotic values of entanglement entropy at $c=1$ and half-filling. The particles are initially only in the left half, and the entropy is computed between the two halves of the chain. For each value of the parameters, $100$ graph samples are taken. The dashed line is the value for a random Gaussian state in the thermodynamic limit. The plot resembles the single-particle behavior.}
    \label{fig:S-vs-alpha-half-filling-c-1}
\end{figure}

\begin{figure}
    \centering
    \includegraphics[width=0.45\linewidth]{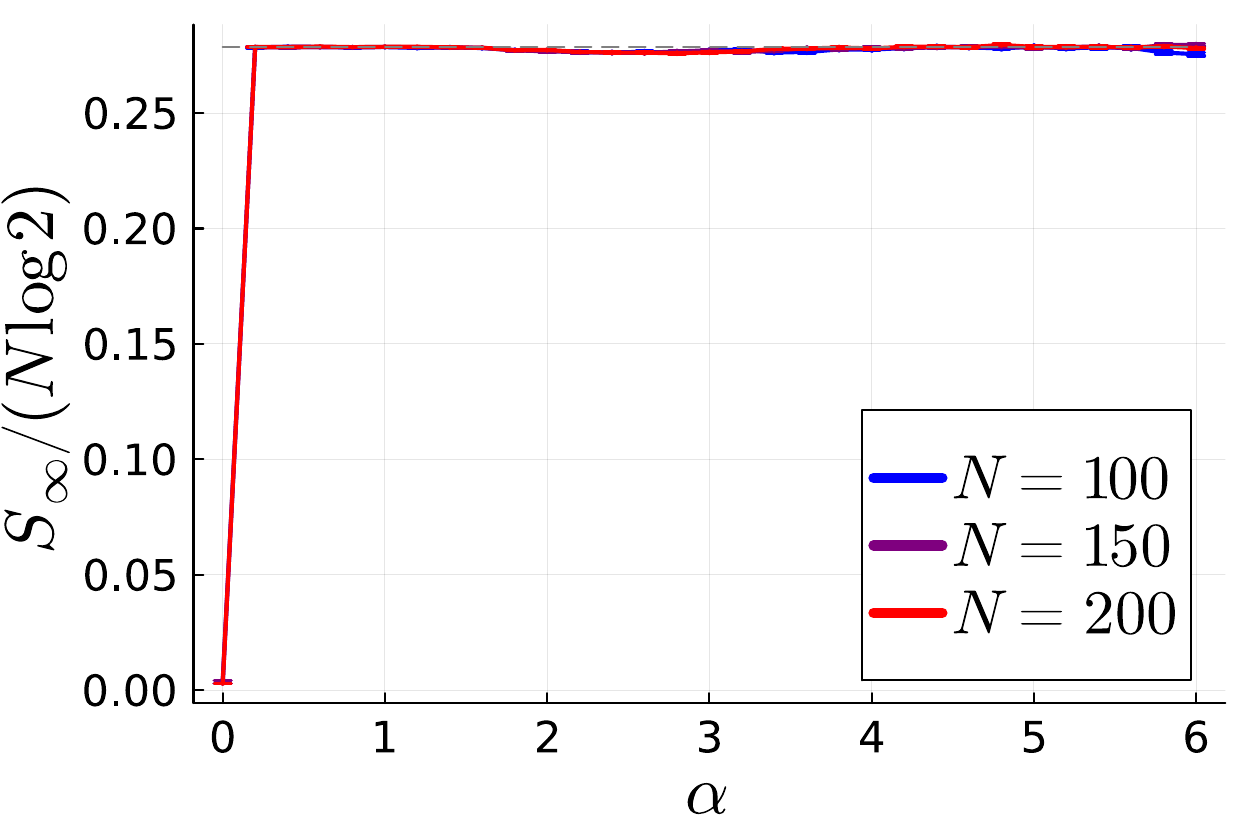}
    \includegraphics[width=0.45\linewidth]{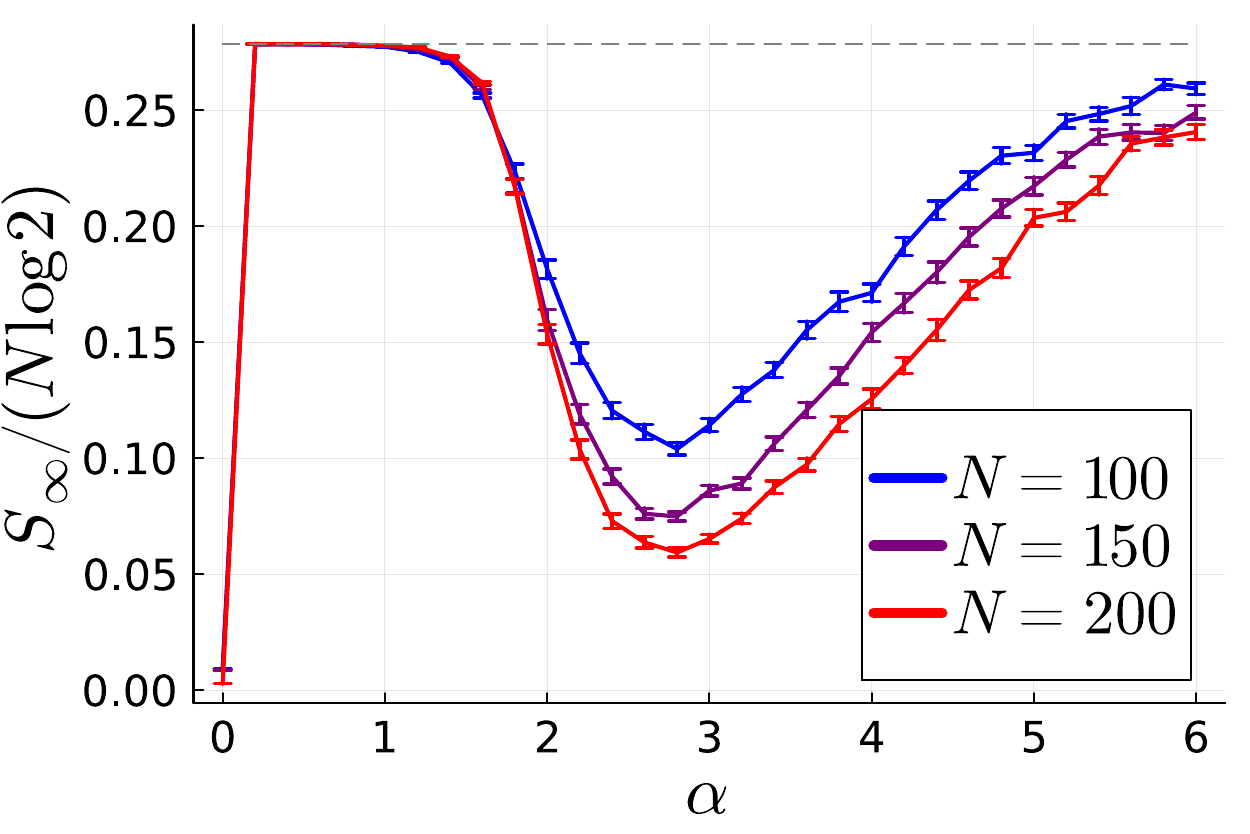}
    \caption{Asymptotic values of entanglement entropy at $c=1$ and half-filling, obtained by initially putting the particles on the odd sites. The bipartition used to compute the entropy is between even and odd sites (left), or between the left and right half of the chain (right). The dashed line is the value for a random Gaussian state in the thermodynamic limit.}
    \label{fig:S-vs-alpha-half-filling-odd-sites}
\end{figure}

We here provide the asymptotic values of entanglement entropy as a function of $\alpha$ at half-filling. In the setup considered, the left half of the chain was initially full and the right half empty, and the entropy is computed with respect to this bipartition of the chain.
The results display similar behavior to the single-particle case, showing the same unexpected localization described in the main text (see \figref{fig:S-vs-alpha-half-filling-c-1}).
Simulations were also performed with the particles initially occupying odd sites, and computing entropy with respect to the even/odd bipartition of the chain, or the left- and right-half bipartition (see \figref{fig:S-vs-alpha-half-filling-odd-sites}). In the first case, the entanglement entropy saturates for each $\alpha \neq 0$ to the value expected for random Gaussian states in the thermodynamic limit. This was to be expected, given the peculiar boundary in this bipartition, which is crossed by all links between nearest-neighbors. Instead, for the other bipartition, we observe the same behavior of the single-particle case. Therefore, the localized regime for intermediate values of $\alpha$ seems to be robust and independent of the initial positions of the particles.

\subsection{Large-$\alpha$ limit at $c=1$}
\begin{figure}[h]
    \centering
    \includegraphics[width=0.5\linewidth]{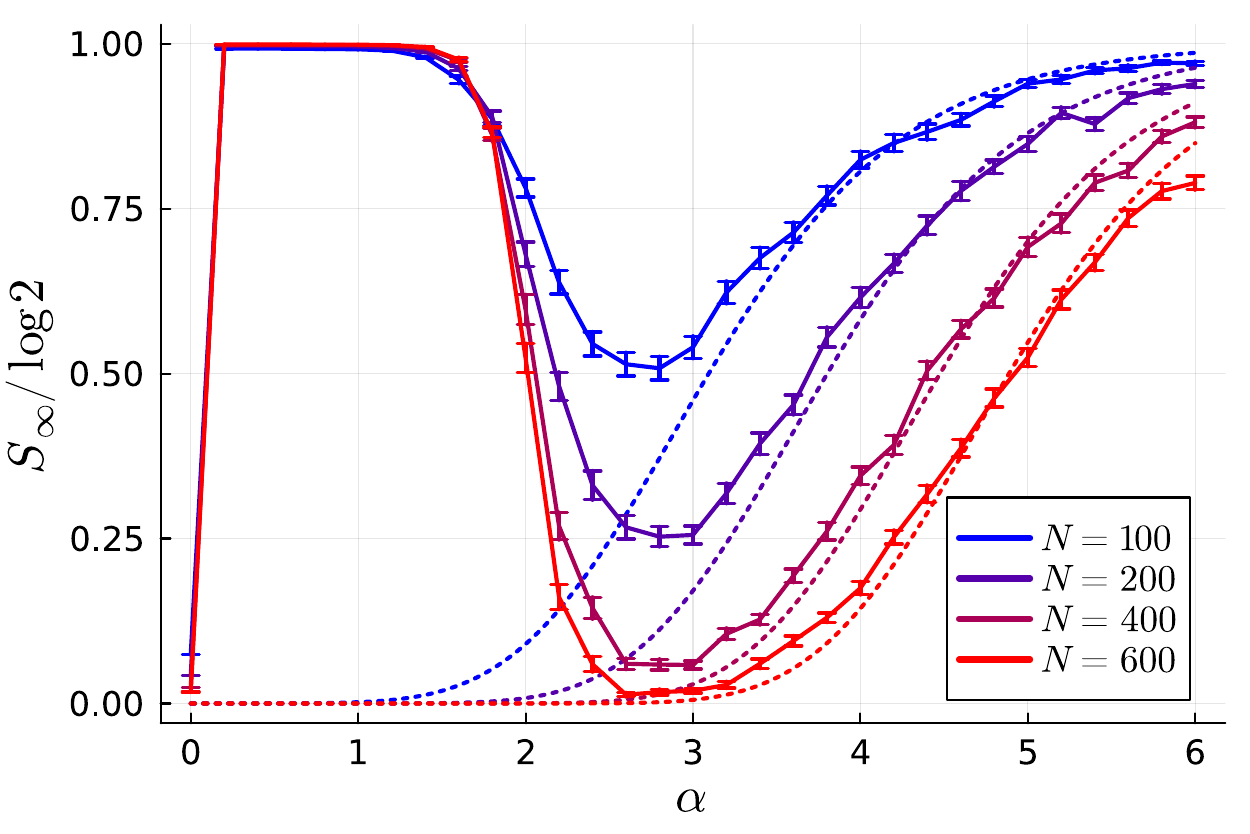}
    \caption{Predictions from the disordered model vs.\ numerical simulations, concerning the large-$\alpha$ behavior of $S_\infty$. Solid lines and corresponding error bars: results of the numerical simulations, see \figref{fig:S-vs-alpha-single-particle}. Dotted lines: best-fit curves from the disordered model, displayed with the same color as the solid lines they refer to. The best-fit curves display approximately the behavior of $S_\infty$ for large $\alpha$, the scatterer model reasonably capturing the underlying physics. As expected, they fail to predict the large values of entropy for $\alpha < 2$. This becomes even more apparent for the smallest values of $N$, for which the minimum entropy is far from zero (a feature that the approximate model is unable to describe). For the values of $N$ shown in the plot, the best-fit values of $A$ are (with increasing $N$) $0.36$, $0.292$, $0.233$, $0.206$.}
    \label{fig:fit-scatterers-model}
\end{figure}

We now focus on the $\alpha \rightarrow +\infty$ limit at $c=1$, in the single-particle case. To build an approximate model in this limit, we neglect all links of length $r > 2$, thus obtaining a chain (from the $r=1$ links) with some additional $r = 2$ links. We then interpret this model as a disordered system in which the $r = 2$ links are impurities, and take $2^{-\alpha}$ as density of the impurities. One-dimensional disordered quantum systems are known to exhibit localization, with the wavefunction amplitude decaying exponentially as the initial particle scatters with the impurities. From this observation, we build an Ansatz for the asymptotic amplitude
\begin{equation*}
    |\psi(x)| \sim \frac{1}{\mathcal N} \exp \left ( -\frac{|x - N/2|}{\lambda} \right )
\end{equation*}
where $\mathcal N$ is a normalization constant and $\lambda$ is the decay length, which can be expressed as
\begin{equation*}
    \frac{1}{\lambda} = 2^{-\alpha} A
\end{equation*}
with $A$ being a constant describing the amplitude absorption from a single impurity. From the Ansatz for $|\psi(x)|$ it is straightforward to compute the asymptotic entanglement entropy $S_\infty$ as a function of $\alpha$ and $N$. We perform a least-squares fit with $A$ as only parameter (see \figref{fig:fit-scatterers-model}).
It should be noted that the decay length $\lambda$ does not depend on $N$, while the size of $\mathcal A$ is equal to $N/2$. For this reason, in this approximate model $S_\infty$ decreases with $N$, in agreement with \figref{fig:S-vs-alpha-single-particle} for large values of $\alpha$.

\subsection{Kac normalization}
We now describe how our results are affected if the Hamiltonian is normalized to be extensive. The effect of Kac normalization~\cite{kac_van_1963} is only a rescaling of the time needed to reach the asymptotic regime, as the normalization factor $\mathcal N$ can be absorbed by redefining $t' = t ~ \mathcal N$.
The average number of terms in the Hamiltonian scales, neglecting factors, like
\begin{equation}
    N \sum_{m = 1}^N m^{-\alpha} .
\end{equation}
For $\alpha > 1$, the summation converges to $\zeta(\alpha)$, with $\zeta$ being the Riemann zeta function. Hence, the normalization factor is simply a constant independent of $N$, and the results remain unaffected. We stress that the unexpected localization and subsequent minimum in the entanglement entropy for intermediate values of $\alpha$ are left untouched by Kac normalization.
For $\alpha \leq 1$, the summation diverges and the normalization is logarithmic in $N$ for $\alpha = 1$, and proportional to $N^{1 - \alpha}$ for $\alpha < 1$.

\begin{figure}[h]
    \centering
    \includegraphics[width=0.5\linewidth]{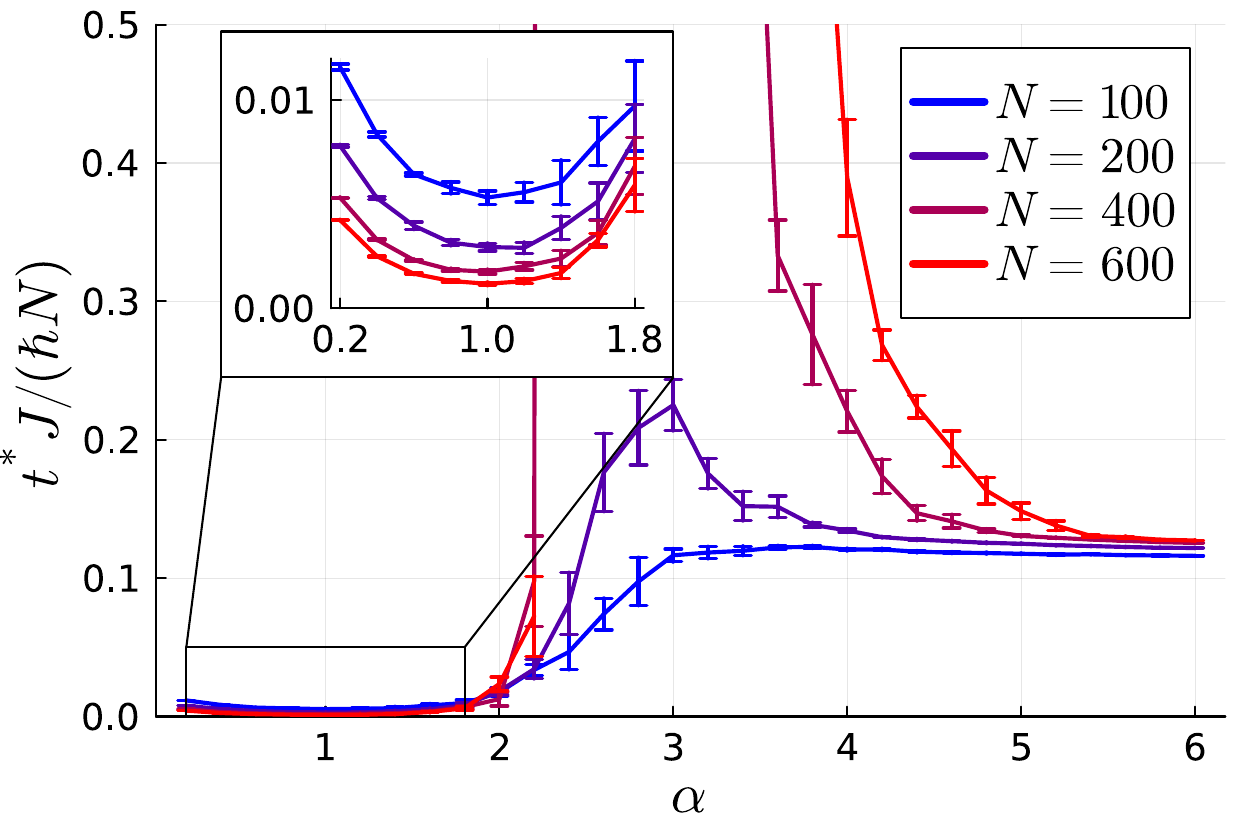}
    \caption{Timescale for the saturation of entanglement entropy to $S_\infty$, after applying Kac rescaling, as a function of $\alpha$. The plot refers to the single-particle case at $c=1$, with a threshold entropy of $S^*/\log 2 = 0.1$. The inset shows that $t^*/N$ decreases with $N$, highlighting that the delocalization for small values of $\alpha$ is still present after Kac-normalizing. As in \figref{fig:timescale}, a divergence in $t^*$ is present for the values of $\alpha$ such that $S_\infty \leq S^*$.}
    \label{fig:timescale-kac}
\end{figure}

For completeness, we also provide a Kac-rescaled plot of $t^*$ as a function of $\alpha$ in \figref{fig:timescale-kac}. It can be noticed that the qualitative behavior is completely unaffected. For $\alpha \in (0,1]$, where the Kac normalization is not a constant, but instead diverges with the system size $N$, one might wonder whether the rescaling will lead $t^*$ to be super-extensive in $N$. The numerical simulations show that this is not the case (see the inset of \figref{fig:timescale-kac}): Hence, the delocalized regime for small values of $\alpha$ may be reached within times sub-extensive in the system size.

\subsection{Features of the spectrum}
\begin{figure}
    \centering
    \includegraphics[width=0.48\linewidth]{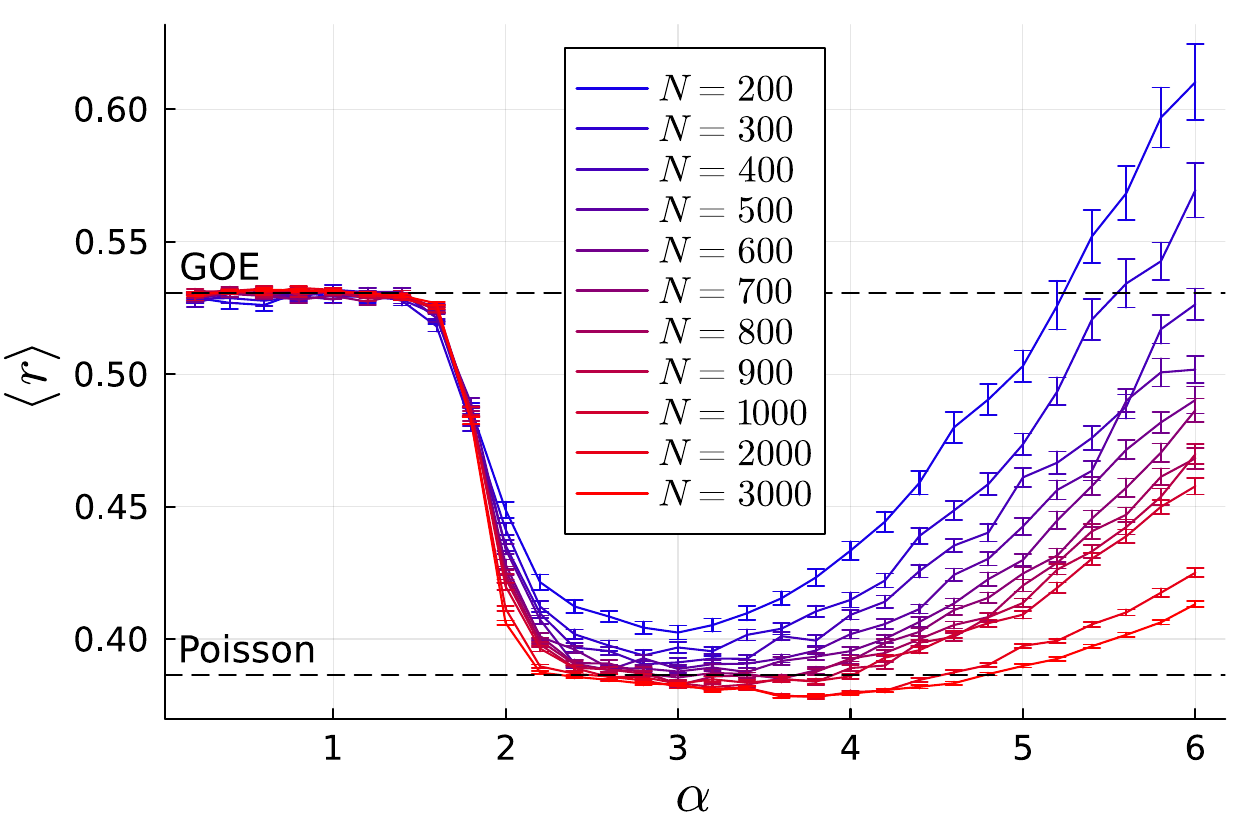}
    \includegraphics[width=0.48\linewidth]{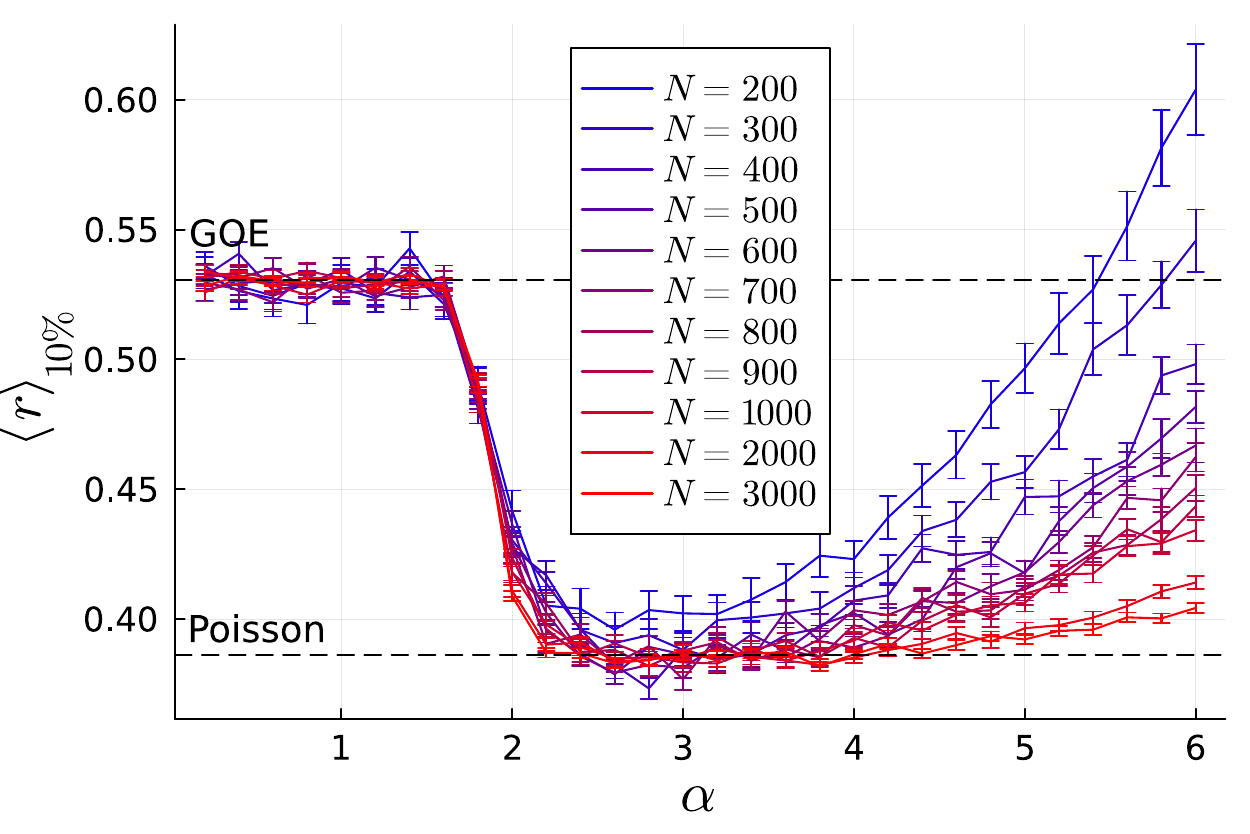}
    \caption{Mean adjacent-gap ratio as a function of $\alpha$. Values are averaged over the whole spectrum ($\langle r \rangle$, left) or the central $10\%$ of eigenvalues ($\langle r \rangle_{10\%}$, right), and then over graph samples. The dashed horizontal lines show the mean ratios for GOE and Poisson distribution.}
    \label{fig:gap-ratio}
\end{figure}

We complement the observations made so far, referring to dynamics, by analyzing spectral properties of the adjacency matrix $A$, and therefore of the Hamiltonian in the single-particle sector.
The first property addressed is the mean adjacent-gap ratio $\langle r \rangle$~\cite{oganesyan_localization_2007,atas_distribution_2013}, defined as the average over the spectrum of
\begin{equation}
    \label{eq:gap-ratio}
    \frac{\min \{ \delta_n, \delta_{n+1} \}}{\max \{ \delta_n, \delta_{n+1} \}} ,
\end{equation}
where $\delta_n$ is the gap between the $n$-th and the $(n+1)$-th eigenvalue, $n = 1, \ldots, N-1$.
The value of $\langle r \rangle$ is then compared to the predictions for Poisson-distributed eigenvalues ($\langle r \rangle = 2 \ln 2 - 1 \approx 0.39$) and for large random matrices in the Gaussian Orthogonal Ensemble (GOE) ($\langle r \rangle \approx 0.53$). The numerical results, obtained by averaging over the full spectrum or only the central $10\%$ of levels, are shown in~\figref{fig:gap-ratio}. Consecutive degeneracies (i.e., occurrences of $\delta_n = \delta_{n+1} = 0$) have been discarded in the computations of the mean ratio, and anyway involved no more than $2\%$ of energy gaps considered for each data point.

In both cases, we observe a crossover between two regimes: one, for $\alpha \lesssim 1.5$, with $\langle r \rangle$ compatible with the GOE value, denoting level repulsion and delocalization; another regime, for $\alpha > 2$, is instead compatible with the value for the Poisson distribution, denoting uncorrelated levels and therefore localization.
The rise in $\langle r \rangle$ for the largest values of $\alpha$ is a finite-size effect, due to the fact that in the large-$\alpha$ limit, at finite $N$, one recovers the clean 1D chain and delocalization.

\begin{figure}
    \centering
    \includegraphics[width=0.48\linewidth]{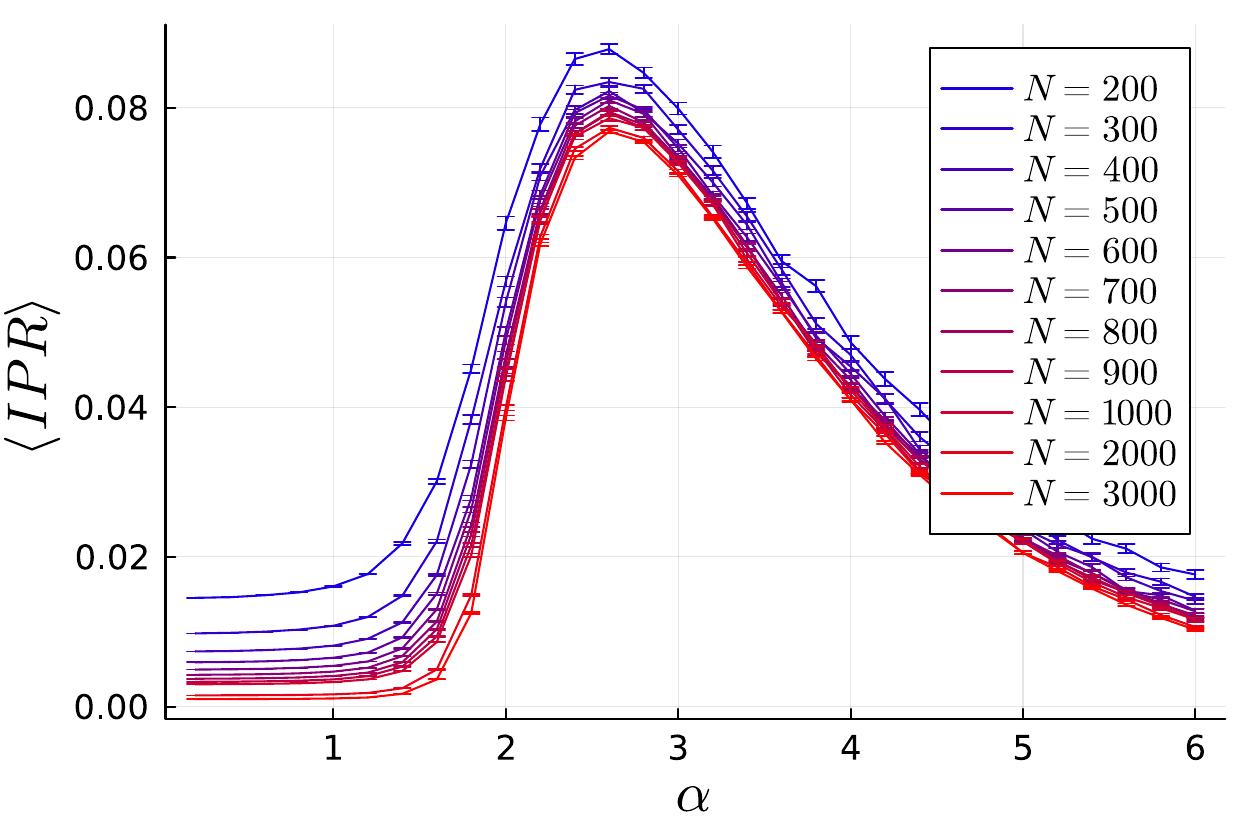}
    \includegraphics[width=0.48\linewidth]{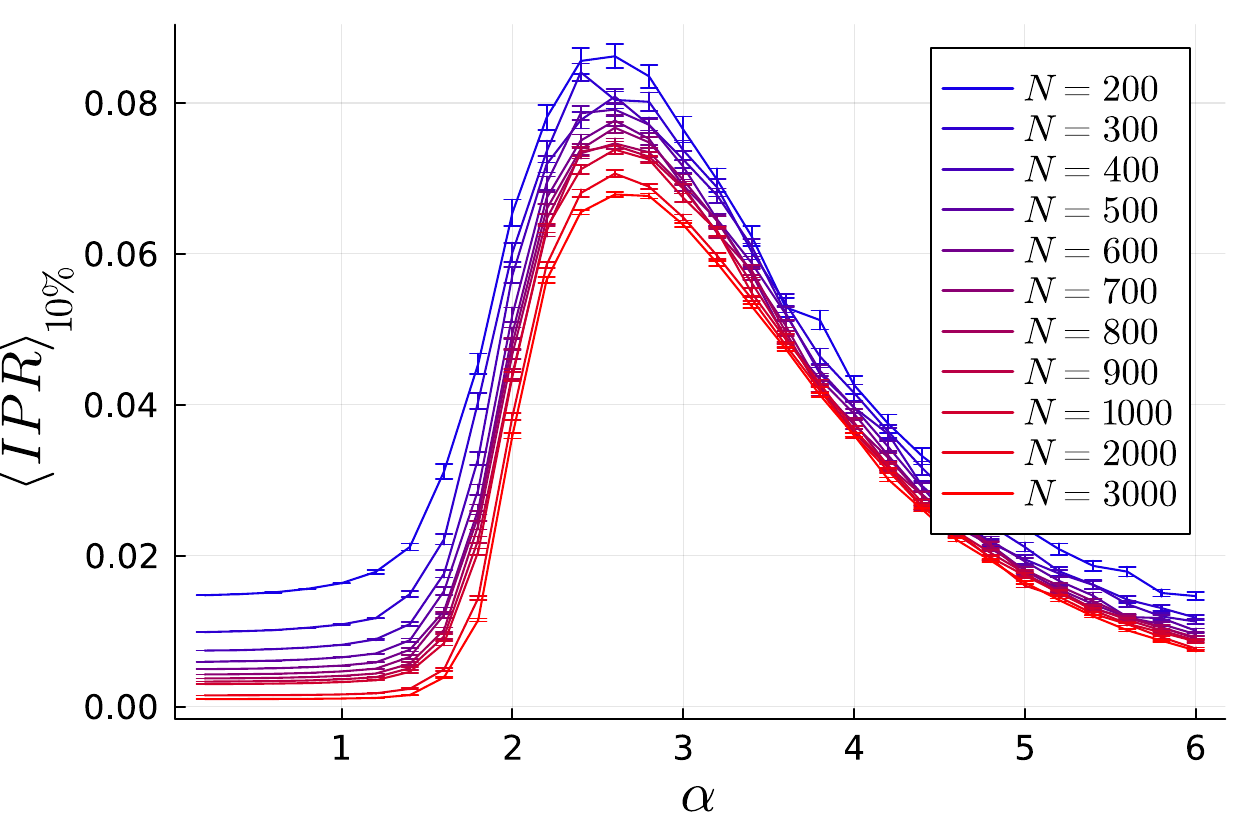}
    \caption{IPR as a function of $\alpha$, averaged over all eigenstates ($\langle \text{IPR} \rangle$, left) or the central $10\%$ of eigenstates ($\langle \text{IPR} \rangle_{10\%}$, right), and then over graph samples.}
    \label{fig:ipr-spectrum}
\end{figure}

\begin{figure}
    \centering
    \includegraphics[width=0.5\linewidth]{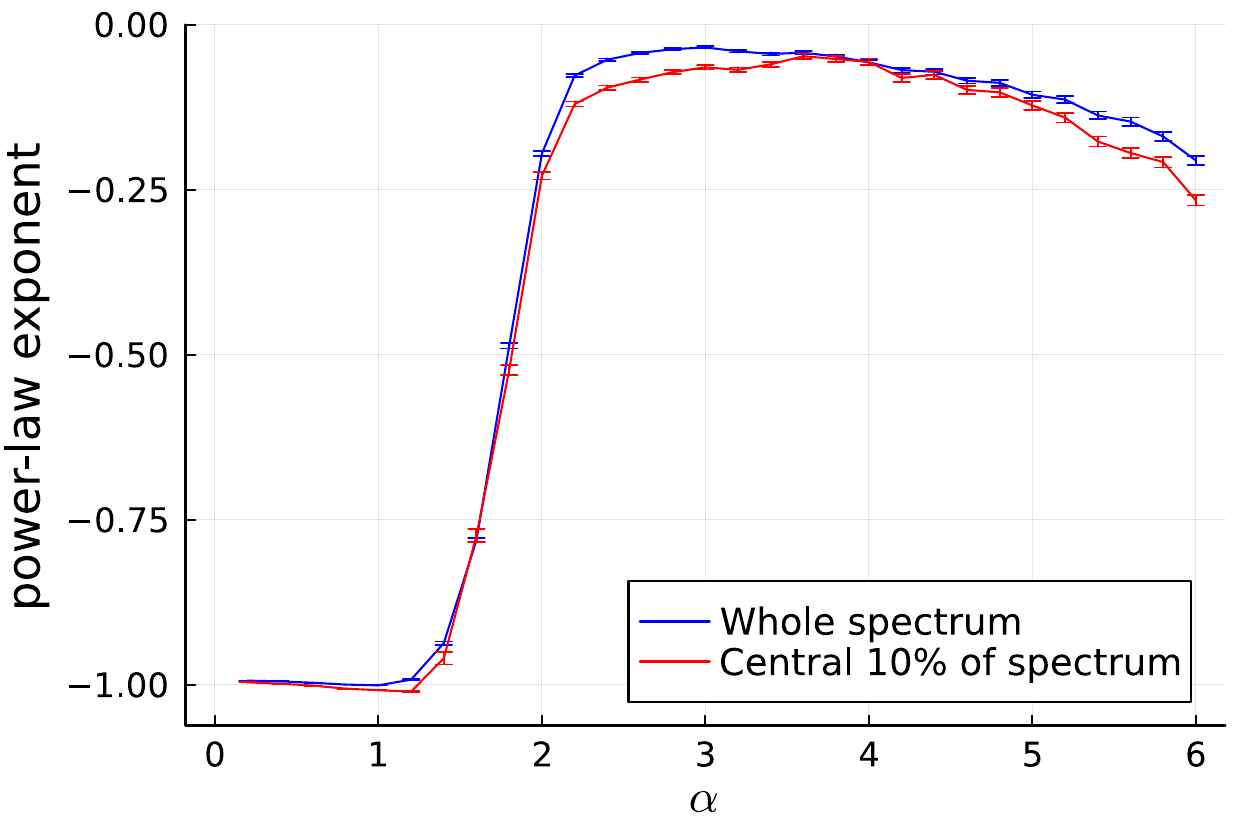}
    \caption{Scaling exponent of the IPR with $N$, as a function of $\alpha$. For each $\alpha$, a linear least-squares fit for $\log IPR = m \log N + q$ is performed, and the slope $m$ is the power-law exponent shown in the plot.}
    \label{fig:ipr-spectrum-scaling}
\end{figure}

Another quantity that has been computed from the spectrum is the average IPR of the eigenstates and its scaling with the system size. Numerical results, again obtained by averaging over the whole spectrum or the central $10\%$ of eigenvalues, can be found in Figures~\ref{fig:ipr-spectrum} and \ref{fig:ipr-spectrum-scaling}.

The scaling of the IPR is typical of delocalization for $0 < \alpha \lesssim 1.5$, with $\text{IPR} \sim N^{-1}$, and therefore a power-law exponent of $-1$. For $\alpha > 2$, the behavior is typical of localization, with a power-law exponent close to zero. The crossover region shows intermediate scalings of the IPR with the system size $N$, a possible hallmark of multifractality.

\end{document}